\documentclass[10pt]{article}
\usepackage{amsmath}
\usepackage{amssymb}
\usepackage{graphicx}

\def\aj{AJ }

\def\mnras{MNRAS }

\begin{document}

\setcounter{figure}{0}
\setcounter{table}{0}
\setcounter{footnote}{0}
\setcounter{equation}{0}

\vspace*{0.5cm}

\noindent {\Large DEVELOPING A PULSAR-BASED TIMESCALE}
\vspace*{0.7cm}

\noindent\hspace*{1.5cm} G. HOBBS$^1$, W. COLES$^2$, R. MANCHESTER$^3$,  
D. CHEN$^4$\\
\noindent\hspace*{1.5cm} $^1$ CSIRO Astronomy and Space Science\\
\noindent\hspace*{1.5cm} PO Box 76, Epping, NSW 1710, Australia\\
\noindent\hspace*{1.5cm} george.hobbs@csiro.au\\
\noindent\hspace*{1.5cm} $^2$ Electrical and Computer Engineering, University of California\\
\noindent\hspace*{1.5cm} La Jolla, California, U.S.A.\\
\noindent\hspace*{1.5cm} bcoles@ucsd.edu\\
\noindent\hspace*{1.5cm} $^3$ CSIRO Astronomy and Space Science\\
\noindent\hspace*{1.5cm} PO Box 76, Epping, NSW 1710, Australia\\
\noindent\hspace*{1.5cm} dick.manchester@csiro.au\\
\noindent\hspace*{1.5cm} $^4$ National Time Service Center\\
\noindent\hspace*{1.5cm} CAS, Xian, China\\
\noindent\hspace*{1.5cm} ding@ntsc.ac.cn\\

\vspace*{0.5cm}

\noindent {\large ABSTRACT.} We show how pulsar observations may be used to construct a time standard that is independent of terrestrial time standards. The pulsar time scale provides a method to determine the stability of terrestrial time standards over years to decades.  Here, we summarise the method, provide initial results and discuss the possibilities and limitations of our pulsar time scale.

\vspace*{1cm}

\noindent {\large 1. INTRODUCTION}

\smallskip

Almost 2000 rapidly rotating neutron stars, known as pulsars, have now been discovered (Manchester et al. 2005)\footnote{http://www.atnf.csiro.au/research/pulsar/psrcat}.  Pulses of radiation from these pulsars are detected using large radio telescopes. The pulses are thought to be caused by a beam of radiation that sweeps across the Earth as the star rotates. As pulsars are very stable rotators, the pulse arrival times (ToAs) can be predicted very accurately over many years.   Recent work (Hobbs, Lyne \& Kramer 2010a)  has shown that, for some pulsars, every rotation of the neutron star over the previous $\sim 40$\,yr can be accounted for.  That work also confirmed the existence of unexplained irregularities in the rotation rate of many pulsars. Such irregularities are generally divided into ``glitch events'' (e.g., Wang et al. 2000) during which the pulsar suddenly increases its spin rate and ``timing noise'' where gradual variations are observed in the pulse arrival times (e.g., Lyne et al. 2010, Shannon \& Cordes 2010). A subset of pulsars, the millisecond pulsars, have rotational periods of a few milliseconds and are extremely stable.  This allows ToAs to be measured with high precision (to within $\sim$50\,ns for some pulsars) and can be accurately predicted over many years using a simple model of the pulsar spin.  

Studies of individual pulsars discovered the first extra-Solar planetary systems (Wolszczan \& Frail 1992), provided tests of the theory of general relativity (Kramer et al. 2006) and have allowed the magnetic field of our Galaxy to be mapped (Han et al. 2006).  In contrast, Foster \& Backer (1990) developed the concept of a ``pulsar timing array" in which many pulsars are observed and phenomena that affect all the pulsar signals are studied.  For instance, the irregular rotation of a pulsar will be uncorrelated with the rotation of a different pulsar.  However, the presence of a gravitational wave passing the Earth will lead to variations in pulse arrival times that are dependent upon the angle between the Earth, pulsar and gravitational-wave source (e.g., Detweiler 1979).  By searching for such correlated signals, an unambiguous detection of gravitational waves may be made.  As described in this paper, irregularities in terrestrial time standards will lead to correlated pulse ToAs (for instance, if the observatory clock is running fast then the pulse arrival times for all pulsars observed will be later than predicted).  

The International Pulsar Timing Array (IPTA) project (Hobbs et al. 2010b) has been developed to search for correlated signals in pulsar data with the main goals of 1) detecting low-frequency gravitational waves, 2) improving the Solar System ephemeris and 3) developing a pulsar-based time scale.  The IPTA project includes data from various observatories.  In this paper we use observations obtained from the Parkes observatory as part of the Parkes Pulsar Timing Array project (PPTA; Verbiest et al. 2010 and references therein).

In \S2 we provide background information relevant to this work.  In \S3 we describe the observations obtained from the PPTA project. In \S4 we overview the method and present our results in \S5.  In \S6 we discuss the implications of a pulsar time scale.  This paper provides an overview; full details of the method and results will be published elsewhere.

\vspace*{0.7cm}

\noindent {\large 2. BACKGROUND}

\smallskip

Atomic frequency standards have been the basis of time keeping since 1955. Individual countries publish local atomic scales.  These are combined by the Bureau International des Poids et Mesures (BIPM) to form International Atomic Time (TAI).  TAI is a realisation of a theoretical time scale known as terrestrial time (TT).  Once published, TAI is never revised, but corrections are published via other realisations of TT.  For instance, TT(BIPM2010) is the most recent post-corrected realisation of TT available.  The difference between this time scale and TAI is shown in Figure 1 from MJD 44000 (the year 1979) to the current date and clearly shows a drift of $\sim 20$\,$\mu$s over $\sim$30\,yr.  As TT(BIPM2010) is a post-corrected version of TT(TAI), it is expected that this drift is caused by inaccuracies in TT(TAI).  However, as TT(BIPM2010) is the world's best time standard, there is no existing method by which a similar figure can be produced to measure inaccuracies in this realisation of TT. 

\begin{figure}[h]
\begin{center}
\includegraphics[width=7cm,angle=-90]{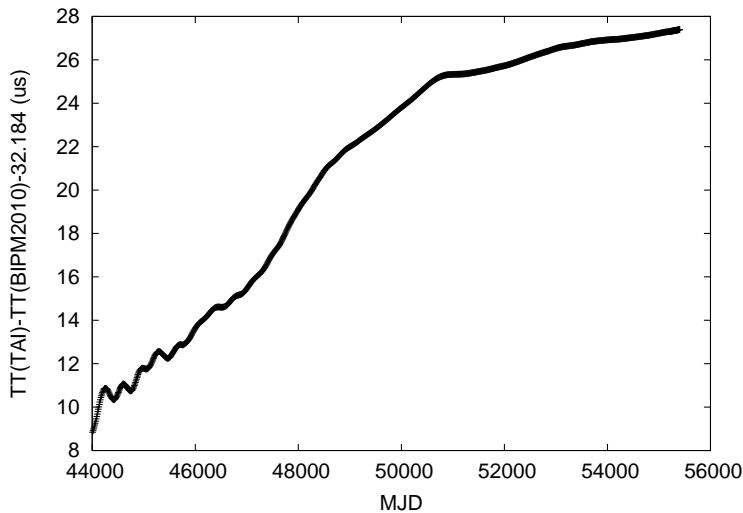}
\caption{The difference between TT(TAI) and TT(BIPM2010).}
\end{center}
\end{figure}

Pulsars, being macroscopic astrophysical objects, are completely unrelated to atomic clocks and are governed largely by unrelated physical principles.   Thus a time standard based upon the rotation of pulsars should be independent of one based on atomic clocks\footnote{Note that the atomic time scale is used in observing pulsars; this ``contradiction'' is discussed in section 6.}.   The rotation of millisecond pulsars is expected to remain stable for billions of years, so they can provide a very long-term standard.  
Matsakis et al. (1997) developed a statistic to compare the stability of pulsar rotation with the stability of various time standards.  A summary of this work is shown in Figure 2 which shows that the two pulsars studied (PSRs B1855$+$09 and B1937$+$21) are significantly less stable than atomic time standards on timescales up to a few years.  On longer timescales the stability of PSR B1937+21 decreases because of timing noise dominating the spin-down of the pulsar.  However, for PSR B1855$+$09 the stability continues to improve and becomes comparable or better than the time standards.

Since the publication of Matsakis et al. (1997) significant improvements have occurred that have allowed many pulsars to be observed with much higher precision.  Such improvements have included new observing instrumentation, the use of higher observing frequencies, the discovery of more stable pulsars and improved calibration procedures.  It is, therefore, now possible to make a pulsar-based time scale that significantly improves on previous attempts.

Various methods have been published describing how to use the pulsar ToAs to form a pulsar time scale (e.g., Guinot \& Petit 1991, Petit \& Tavella 1996, Foster \& Backer 1990 and Rodin 2008).   The initial stage of all these methods is to form ``pulsar timing residuals'' which are the difference between the measured arrival times and predictions of these times given a  model of the pulsar (see, e.g., Edwards, Hobbs \& Manchester 2006). If the model for the pulsar is ``perfect'' then there will be no statistically significant timing residuals.  However, if the Earth-based time scale is incorrect then the pulse arrival times measured using this time scale will deviate from the predicted arrival times resulting in significant timing residuals.  Previous pulsar time scales all involved taking a suitably weighted linear combination of the timing residuals from the various pulsars, exactly as one determines a time scale from an array of atomic clocks. However these methods are not optimal for pulsars because they do not take into account the irregular nature of pulsar observations and the interaction between fitting the timing model and taking a weighted sum of the residuals. In fact the problem is best expressed statistically as recovering a signal which all the ToAs have in common.

\begin{figure}[h]
\begin{center}
\includegraphics[width=7cm]{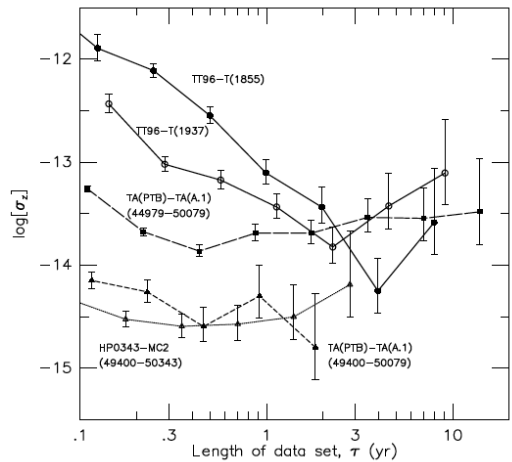}
\caption{Figure taken from Matsakis et al. (1997) comparing the stability of PSRs B1855+09 and B1937+21  with various atomic clock differences.}
\end{center}
\end{figure}

\vspace*{0.7cm}
\noindent {\large 3. OBSERVATIONS}

\smallskip

Here we make use of observations from the Parkes telescope. These observations, of 19 millisecond pulsars, were presented in Verbiest et al. (2008) and Verbiest et al. (2009).  For this paper we do not make use of PSR~J1939$+$2134, which was also published in Verbiest et al. (2009) as its timing residuals are dominated by timing noise and our algorithm gives it essentially zero weight. All observations were carried out in the 20\,cm band except for PSR~J0613$-$0200 for which 50\,cm observations have been used.  Most of the processing was identical to that described in Verbiest et al. (2008/2009).  However, we have now measured the majority of the timing offsets between the different instruments.  Timing residuals were formed using the \textsc{tempo2} software package (Hobbs, Edwards \& Manchester 2006) using the JPL DE414 ephemeris and referred to TT(TAI).   

\begin{figure}[h]
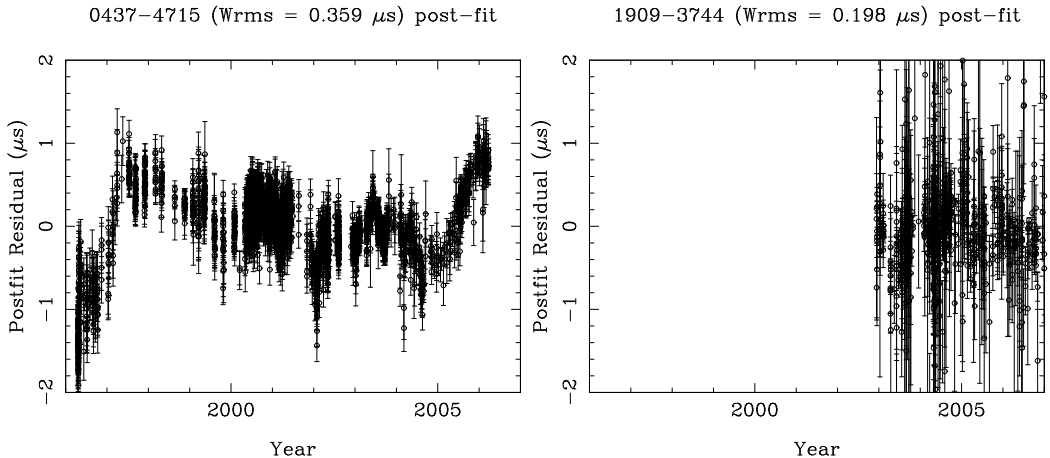

\begin{center}
\includegraphics[width=6cm,angle=-90]{hobbs_03.ps}
\includegraphics[width=6cm,angle=-90]{hobbs_04.ps}
\caption{Timing residuals for PSRs~J0437$-$4715 (left) and J1909$-$3744 (right) displayed with the same axis scaling.}
\end{center}
\end{figure}

The timing residuals for two pulsars, PSRs~J0437$-$4715 and J1909$-$3744 are shown in Figure 3.  This figure highlights various issues that need to be considered when developing a pulsar time scale:
\begin{itemize}
\item{The observations of different pulsars cover different ranges and do not necessarily overlap.}
\item{The ToA uncertainties vary by several orders of magnitude in time because of different observing durations, scintillation and improvements in observing systems.  The precision with which ToAs can be determined can vary by many orders of magnitude between different pulsars.}
\item{The observations are unevenly sampled and large gaps (of many months or years) exist for some pulsars.}
\item{The pulsar timing model is formed by least-square-fitting to the timing residuals.  This process will always fit and remove a quadratic polynomial from the timing residuals.}
\item{Unmodelled timing noise is apparent for some pulsars. It is uncorrelated between pulsars.}
\end{itemize}
We emphasise that the fit and subsequent removal of the quadratic polynomial implies that we can never detect linear or quadratic drifts in terrestrial time standards (the long-term drift in Figure 1 could not be detected using pulsars, however the smaller, faster variations are detectable).

\vspace*{0.7cm}
\noindent {\large 4. METHOD}

\smallskip
We have developed a method for comparing the timing residuals from multiple pulsars and determining the correlated signal. Details of the method and the tests undertaken to confirm its validity will be described in a separate paper.  Here we provide an overview.  

We choose to model the correlated signal (i.e., the clock error) as a Fourier series to provide some control of the covariance between the quadratic in the timing model and the clock error:
\begin{equation}
x_c(t) = \sum_{k=1}^{n} A_k \cos(k\omega_0t) + B_k\sin(k\omega_0t)
\end{equation}
where $2\pi/\omega_0$ is chosen to be close to the total span of the observations.  The total number of harmonics $n$ is dependent upon the exact data being analysed.  The parameters $A_k$ and $B_k$ are fitted as part of the standard \textsc{tempo2} procedure which has been updated to include multiple pulsars simultaneously.  We whiten and normalise the residuals for each pulsar using the Cholesky method which has recently been implemented in \textsc{tempo2} (Coles et al., submitted to MNRAS).  The fitting process provides the covariance matrix for the parameters $A_k$ and $B_k$ from which we calculate the uncertainty of $x_c(t)$. The resulting correlated signal is referred to TT(TAI) so the correlated signal, $x_c(t)$, can be regarded as TT(PSR)-TT(TAI).

\vspace*{0.7cm}
\noindent {\large 5. RESULTS}

\smallskip

\begin{figure}[h]
\begin{center}
\includegraphics[width=7cm,angle=-90]{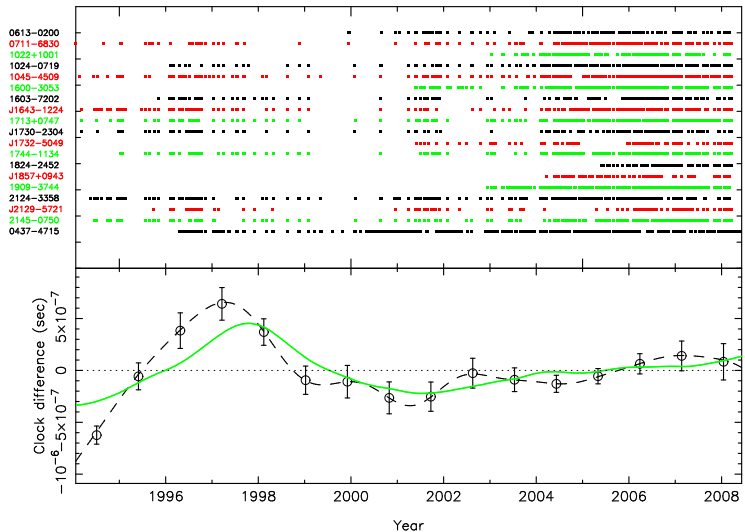}
\caption{The top panel shows the data sampling for the different pulsars in the sample.  The bottom panel shows the pulsar time scale with respect to TT(TAI).  The solid line indicates TT(TAI)-TT(BIPM2010) with a quadratic polynomial fitted and removed.}
\end{center}
\end{figure}

The sampling and data spans for the 19 pulsars included in this analysis are shown in the top panel of Figure 4.  The dashed line connecting the error bars in the bottom panel is $x_c(t)$ and the solid line is TT(BIPM2010)-TT(TAI) with a best-fit quadratic removed. If the observations were equally spaced and the observational errors were equal our procedure would be equivalent to a spectral analysis and the error bars would be independent. In practice the error bars are correlated, but we have spaced them such that the correlation is not too high.  It is clear that the pulsar timescale is very similar to
TT(BIPM2010); they show similar deviations from TT(TAI) around 1997 and a similar linear trend more recently. There is a marginally significant difference between $x_c(t)$ and TT(BIPM2010) between 1996 and 1998.

\vspace*{0.7cm}
\noindent {\large 6. DISCUSSION AND CONCLUSIONS}

\smallskip

Here we outline and summarise the discussion that followed the presentation of this paper.

\begin{itemize}
\item \emph{Can any phenomena mimic a terrestrial time standard error?} We have detected significant correlations in the pulsar timing residuals and have attributed these correlations to the terrestrial time standard.  Errors in the time transfer to the Parkes observatory would appear in $x_c(t)$, but these are thought to be $<$20\,ns.  The possible effects of gravitational waves are also partially correlated, but these are expected to be $<$100\,ns (Yardley et al., in preparation). 

\item \emph{Can a weighted average approach be used to form the pulsar based time standard?} We do not believe that a weighted average approach is a reliable technique.  Step errors will occur at the beginning and end of shorter data sets and these will require smoothing.  Some of the variations will be spuriously reduced by the independent quadratic removal.

\item \emph{Is the pulsar time scale independent of atomic clocks and is the pulsar time a realisation of terrestrial time?} In order to form pulsar timing residuals it is necessary that a precise time is recorded with each observation and that this time can be converted to Barycentric Coordinated Time (TCB) via a realisation of terrestrial time.  However, assuming that both the conversion from TT to TCB and the transfer from the observatory time to TT are known to within a few nanoseconds then any induced correlated timing residuals will be caused by irregularities in the realisation of TT.  Subtracting these correlated residuals from the realisation of TT produces a new, improved realisation, TT(PSR). TT(PSR) is presented in the form of corrections to TT(TAI) exactly as is TT(BIPM2010).

\item \emph{What is the effect of correcting TAI?} The difference between TT(TAI) and TT(BIPM2010), shown in Figure~1, is a result of a deliberate correction of the primary standard frequencies.  Around 1993 the frequency of TAI was found to be too low (Petit 2004) which resulted in the decision in 1995 to correct the primary frequency.  This correction was phased in over three years (Petit, 2004) which resulting in the obvious ``bump'' in Figure 1.   Around the year 2000, when caesium fountains began to contribute significantly to TAI, further steering has been necessary. These corrections should be reflected in TT(BIPM2010) so it should agree with TT(PSR).  However, it is possible that the corrections in TT(BIPM2010) were not perfect and some of the difference between TT(BIPM2010) and the pulsar time scale may be real.

\end{itemize}

Pulsar timing precision is continuing to improve as new pulsars are discovered, new telescopes built and new instrumentation installed.  Combining data from observatories world-wide will lead to a large increase in the number of pulsars and the data span compared with our PPTA observations.  Comparing data from different observatories also provides a means by which irregularities in the time transfer from the observatory clock to TT can be determined and removed.  Within a few decades it is likely that the Square Kilometre Array (SKA) telescope will revolutionise pulsar timing.  We therefore believe that pulsar timing will continue to advance in step with the expected improvements in atomic clocks and continue to provide a valuable independent time reference. 

\vspace*{0.7cm}

\noindent {\large 7. ACKNOWLEDGEMENTS}

\smallskip

 The Parkes Pulsar Timing Array project is a collaboration between numerous institutes in Australia and overseas and we thank our collaborators on this project.  The Parkes radio telescope is part of the Australia Telescope which is funded by the Commonwealth of Australia for operation as a National Facility managed by CSIRO.  GH acknowledges support from the Chinese Academy of Sciences \#CAS KJCX2-YW-T09, NSFC 10803006 and from the Australian Research Council (project \#DP0878388).

\vspace*{0.7cm}
\noindent {\large 8. REFERENCES}
% Please type the reference as follows
% Name Initial, year, "title", journal, vol. , pp. x-x.
%
% Examples:
%

% Author1, N., Author2, N., 2000, ``Title of the paper'', 
% \aa 111, pp. 111--222.
%
% Author2, N., Author3, N., 2003, ``Title of the paper'',
% \jgr (Solid Earth), 111(B5), doi: 10.1000/2002JB001111.
%
% PLEASE DO NOT USE ANY SPECIAL FONTS 
% (no italics, no boldface, etc.)
%
{

\leftskip=5mm
\parindent=-5mm

\smallskip

Detweiler, S., 1979, ``Pulsar timing measurements and the search for gravitational waves'', ApJ, 234, 1100

Edwards, R., Hobbs. G, Manchester, R. N., 2006, ``TEMPO2, a new pulsar timing package - II. The timing model and precision estimates'', MNRAS, 372, 1549

Foster, R. S., Backer, D., C., 1990, ``Constructing a pulsar timing array'', ApJ, 361, 300

Guinot, B., Petit, G., 1191, ``Atomic time and the rotation of pulsars'', A\&A, 248, 292

Han, J. L., et al., 2006, ``Pulsar Rotation Measures and the Large-Scale Structure of the Galactic Magnetic Field'', ApJ, 642, 868

Hobbs, G., Lyne, A. G., Kramer., M., 2010a, ``An analysis of the timing irregularities for 366 pulsars'', \mnras, 402, 1027

Hobbs, G., et al., 2010b, ``The International Pulsar Timing Array project: using pulsars as a gravitational wave detector'', CQGra, 27, 4013

Hobbs, G., Edwards, R. T., Manchester, R. N., 2006, ``TEMPO, a new pulsar-timing package - I. An overview'', MNRAS, 369, 655

Kramer, M., et al. 2006, ``Tests of General Relativity from Timing the Double Pulsar'', Science, 314, 97

Lyne et al., 2010, ``Switched Magnetospheric Regulation of Pulsar Spin-Down'', Science, 329, 408

Matsakis, D. N., Taylor, J. H., Eubanks, T. M., 1997, ``A statistic for describing pulsar and clock stabilities'', A\&A, 326, 924

Manchester, R. N., Hobbs, G. B., Teoh, A., Hobbs, M., 2005, ``The Australia Telescope National Facility Pulsar Catalogue'', \aj 129, 1993

Petit, G., 2004, www.bipm.org/cc/CCTF/Allowed/16/cctf04-17.pdf

Petit, G., Tavella, P., 1996, ``Pulsars and time scales'', A\&A, 308, 290

Rodin, A. E., 2008, ``Optimal filters for the construction of the ensemble pulsar time'', MNRAS, 387, 1583

Shannon, R., Cordes, J. M., 2010, ``Assessing the Role of Spin Noise in the Precision Timing of Millisecond Pulsars", arXiV, 1010, 4794

Verbiest, J. P. W. et al., 2010, ``Status update of the Parkes pulsar timing array'', CQGra, 27, 4015

Verbiest, J. P. W. et al. 2009, ``Timing stability of millisecond pulsars and prospects for gravitational-wave detection'', MNRAS, 400, 951

Verbiest, J. P. W. et al. 2008, ``Precision timing of PSR J0437-4715: An accurate pulsar distance, a high pulsar mass, and a limit on the variation of Newton's Gravitational constant'', ApJ, 679, 675

Wang et al., 2000 ``Glitches in Southern Pulsars", MNRAS, 317, 843

Wolszczan, A., Frail, D. A., 1992, ``A planetary system around the millisecond pulsar PSR1257+12'', Nature, 355, 145

}

\end{document}